# PIC simulation of current-driven solar type III radiation and whistler waves in an electron core-strahl plasma: Relevance to PSP and other space observations

Konrad Sauer[1] and Kaijun Liu[2]

[1] *Max-Planck Institute of Solar System Research, Göttingen, Germany*
[2] *Southern University of Science and Technology, Shenzhen, Guangdong, China*

**Abstract** The aim of the paper is to demonstrate that electron current oscillations may generate electromagnetic waves as type III radiation and whistler waves without the involvement of the classical plasma emission via the coalescence of waves. PIC simulation results of an electron-core-strahl plasma without initial current compensation are presented which describe the conversion of current-driven Langmuir oscillations/waves into type III radiation whereby simultaneously whistler waves are excited. In contrast to the classical approach of Ginzburg and Zhelezniakov (1958) after which beam-excited Langmuir waves in a two-step process are converted in electromagnetic radiation, any instability is suppressed by selecting a low strahl velocity. Rather electric field oscillations at the electron plasma frequency are triggered by the initially non-compensated current of the strahl. The arising electromagnetic fields exhibit amplitude oscillations which are caused by the superposition of the two wave modes of mixed polarisation at the point of mode coupling. This basic mechanism of wave generation and transformation has already been described in earlier papers using simple fluid models. It is also the topic of the companion paper. Besides the fundamental electromagnetic radiation, the second harmonic of nearly the same intensity has been obtained which is an indication for nonlinear currents. Measurements of Langmuir waves, type III radiation and whistler waves on board various satellites in the solar wind, in particular Parker Solar Probe (PSP) observations are analysed in the light of our results. Interpretations of earlier PIC simulations are critically reviewed.

## 1. Introduction

The simultaneous detection of Langmuir and ion-acoustic waves in the occurrence of type III radio bursts since the early space missions (e.g. Gurnett and Anderson, 1976) has led to the two-stage process of their generation mechanism developed by Ginzburg and Zhelezniakov (1958) being accepted as generally valid to this day. Although some problems have become apparent, such as the Sturrock (1964) paradox on the contradiction between the short-term beam instability and the observed long-lived type III radio bursts, it is generally assumed that Langmuir waves excited by electron beams form the starting point for subsequent non-linear wave-wave and wave-particle interactions of conversion into electromagnetic radiation. In this sense, radio waves have been considered as a powerful tool for the diagnostics of electron beams. Because of the extensive literature on this problem, reference is made here to reviews, as Reid and Ratcliffe (2014). A similar situation exists in theoretical studies (Kasaba et al., 2001; Rhee et al., 2009; Henri et al., 2019; Zhang et al., 2022; Chen et al., 2022) where 1D and 2D particle-in-cell (PIC) simulations have been used. Here, high-energetic electron beams are assumed at t = 0 and the detection of (forward and backscattered) Langmuir and ion-acoustic waves is taken as evidence that the obtained electromagnetic radiation at the plasma frequency and its second harmonic has its origin in the classical mechanism of plasma emission according to Ginzburg and Zhelezniakov (1958).

However, doubts about the traditional model of plasma emission arose some time ago. They go back to first studies in Sauer and Sydora (2012) in which the effects of mode coupling have been investigated by means of PIC simulations. Using the 1D electromagnetic code, the simulation of the classical beam-instability has shown that besides growing Langmuir waves in the range of resonant wave numbers k given by $\omega_e = kV_b$ ($\omega_e$: electron plasma frequency, $V_b$: beam velocity), an excitation of the left-handed electromagnetic (L) wave takes place where the Langmuir wave and the L wave (also called z mode) cross each other. That is at wave numbers of $kc/\omega_e < 1$. Detailed analysis brought about the conclusion that this effect cannot be reconciled with the two-step model of Ginzburg and Zheleszniakov (1958). Simultaneously, whistler waves were excited which could not be directly connected to the beam due to a lack of resonance.

A new element came into the considerations with the concept of (current-driven) Langmuir waves (Sauer and Sydora, 2015), which are not driven by an instability, but by an initial current. Later, the fluid approach has been extended by taking into account electromagnetic effects (Sauer et al., 2019) in which the coupling between the current-driven Langmuir wave and the adjacent L-wave is contained. Stimulated by recent PSP measurements, e.g., by Agapitov et al. (2020), our attention has also been focused on the excitation of low-frequency waves. In the end, it turned out that simultaneous excitation of high-frequency waves near $\omega_e$ and of whistler waves with $\omega \leq 0.1\Omega_e$ is an essential element of the theory of current-driven waves. Indeed, the observation of low-frequency whistler waves related to Langmuir waves can be found in the literature, e.g., in Moullard et al. (1998, 2001). It has recently aroused new interest through measurements on PSP concerning their origin, see, e.g., Jagarlamudi et al. (2021).

As pointed out already, the simultaneous excitation of electromagnetic waves in different frequency ranges by an initial current is an essential output of the fluid theory which is described in the companion paper by Sauer and Liu (2025). However, the crucial question that remained unanswered was to what extent the predictions made there could be confirmed by the necessary inclusion of kinetic effects. This was the starting point for particle simulations, which were carried out in parallel with the development of the fluid model. The main considerations were to select the parameters for the kinetic simulations in such a way that the formation of Langmuir waves is suppressed. For this purpose, a plasma was chosen whose electron distribution is composed of a core and a strahl, whereby the strahl velocity was chosen low enough to prevent beam instability. Recent measurements by Mozer et al. (2024) aboard PSP were used for orientation and accordingly a strahl density of 5% of the core density and a strahl velocity of four times the thermal velocity of the core were selected in the simulations.

Some remarks should be made in advance to clarify the differences between our approach and previous PIC simulations. Their main focus was to verify the plasma emission process according to the model of Ginzburg and Zheleznikov (1958). Accordingly, almost exclusively 2D models were used to include the possibility of omni-directional wave-wave interaction, e.g., Kasaba et al. (2001), Rhee et al. (2009), Thurgood and Tsiklauri (2015), Henri et al. (2019), Lee et al. (2019), Zhang et al. (2022), Chen et al. (2022), and Krafft et al. (2024). 1D electromagnetic PIC simulations to study beam-plasma interaction have only been used in the paper by Sauer and Sydora (2012) whose results, as mentioned already, raised initial doubts about the classical concept of plasma emission, whereby the assumption of oblique propagation to the background magnetic field was an important point.

Preliminary considerations for triggering Langmuir oscillations/waves are important of the further understanding. From the very beginning, it was an important question for us how to convincingly represent the mechanism of current-driven Langmuir oscillations/waves and the associated radiation generation. If one assumes a beam instability, as in previous simulations, one faces the problem that the processes associated with the instability of classical plasma emission (generation of Langmuir waves with $k\lambda_D < 1$ ($\lambda_D$: Debye length), parametrically generated ion waves, backscattering, etc.) mask the direct excitation of plasma waves with $kc/\omega_e \sim 1$. In addition, the wavenumber resolution in 2D PIC simulations is generally insufficient for this. Therefore, our effort was directed towards using an initial configuration in which current-driven Langmuir waves play a role without a beam instability being involved. In PIC codes, this combination can be most easily realized without changes to the existing routines by assuming a sudden onset of a strahl whose current is not compensated by a corresponding shift in the core distribution. This point also touches on the controversial discussion in literature regarding the consequences of an initial net current induced by the beam if no compensation is made by a corresponding shift of the principal electron component. While the 2D simulations by Thurgood and Tsiklauri (2015) link the detected Langmuir oscillations at $k \sim 0$ and the electromagnetic second harmonic to the non-zero current condition, Henri et al. (2019) see no decisive influence of the net current. These uncertainties result from the insufficient resolution of 2D simulations in the range of wavenumbers which mostly exceeds $kc/\omega_e \geq 1$. The study of the conversion of Langmuir oscillations/waves into electromagnetic radiation, however, requires a much better resolution which has not been feasible so far.

This is where our considerations come in. These were aimed at taking advantage of a spatially one-dimensional simulation with high resolution in the wavenumber range to focus exclusively on the processes of wave conversion at very low wavenumbers ($kc/\omega_e < 1$). To capture only the ignition of waves by an electron current at $t = 0$, an electron distribution consisting of a core and a strahl was chosen that does not allow for beam instability. Any coalescence of waves is completely outside our scope. Importantly, however, the oblique propagation of the waves with respect to the external magnetic field is assumed. Only then is there a coupling between the Langmuir and the left-circular electromagnetic (L) waves, which is ultimately crucial for the generation of type III radiation. To summarize, one can say that 1D simulations considering oblique wave propagation with respect to the background magnetic field represent a sufficient tool of the envisaged studies.

The outline of our paper is the following: In Section 2, a 1D PIC code is used to simulate the electromagnetic wave generation in a plasma consisting of core and strahl electrons. The corresponding wave dispersion from the Vlasov approach is discussed in Subsection 2.1 whereby the wave propagation direction with respect to the background magnetic field is restricted to quasi-parallel. A short description of the 1D PIC code is given in 2.2. The arising oscillations of both electron populations which are driven by the initial current of the strahl are shown in 2.3. As a main part of the paper, Section 2.4 is devoted to the conversion of the current-driven Langmuir oscillations in type III radiation and to the generation of whistler waves. Similarly, in Section 3, the PIC simulation is applied to the case where the initial current is caused by a drift of the electrons with respect to the ions. With respect to the generation of electromagnetic radiation, the results are completely comparable. Furthermore, significant changes in the electron distribution function are discussed. The occurring oscillations of the electrons lead to scattering processes that result in the heating of the tail and the formation of a halo distribution. In Summary and Discussion, relevant space observations are discussed in the light of the theoretical results. Comments on the interpretation of previous work on the mechanism of plasma emisssion are included.

## 2. Electromagnetic wave generation in a core-strahl plasma

In the subsequent PIC simulations, an initial plasma configuration close to real solar wind conditions, consisting of core and strahl electrons, is assumed. With respect to the relevant parameters, the measurements presented by Mozer et al. (2024) serve as an orientation. Accordingly, for the density and velocity of the strahl (s), the same parameters as in the fluid approach of the companion paper (Sauer and Liu, 2025) are taken: $n_s/n_0 = 0.05$ and $V_s = 4V_e$. Here, $n_0$ is the total electron density, $V_e$ stands for the thermal velocity of the core whereby the same is taken for the strahl.

### 2.1 Wave dispersion of a core-strahl plasma:

Figure 1 shows the initial electron distribution function and the dispersion of related waves in the wave number range up to $kc/\omega_e = 10$ for the propagation angle $\theta = 5^0$ and $G = \Omega_e/\omega_e = 0.1$, where c is the speed of light and $\Omega_e$ is the electron cyclotron frequency. In addition to the wave modes of a pure core plasma, the presence of the strahl leads to the occurrence of an additional electron-acoustic mode (called the strahl mode) which intersects the Langmuir mode at $kc/\omega_e \sim 6$. Ordering the wave modes at $kc/\omega_e = 1$ from high to low frequencies, one has the following: The electromagnetic R and L modes ($\omega \sim 1.4\omega_e$), the Langmuir wave ($\omega \sim \omega_e$), the strahl mode, and the whistler mode ($\omega \leq 0.1\omega_e$). At the point of intersection of the strahl and Langmuir modes ($kc/\omega_e \sim 6$), a weak beam instability appears. As shown later, however, this has no influence on the conversion of the current-driven Langmuir oscillations into electromagnetic radiation. After a very short time, the bump-on tail configuration turns into a stable plateau. More important for the process of electromagnetic wave generation is the mode coupling, which takes place in the range of small wave numbers ($kc/\omega_e < 1$) between the Langmuir wave and the electromagnetic L-wave. The dispersion of these waves together with the whistler wave is therefore highlighted once again in Figure 2 for two values of G taking a propagation angle of $\theta = 15^0$. As discussed in the companion paper, the corresponding wave number of the mode crossing point shifts with the parameter G according to $kc/\omega_e \sim [G/(1+G)]^{1/2}$.

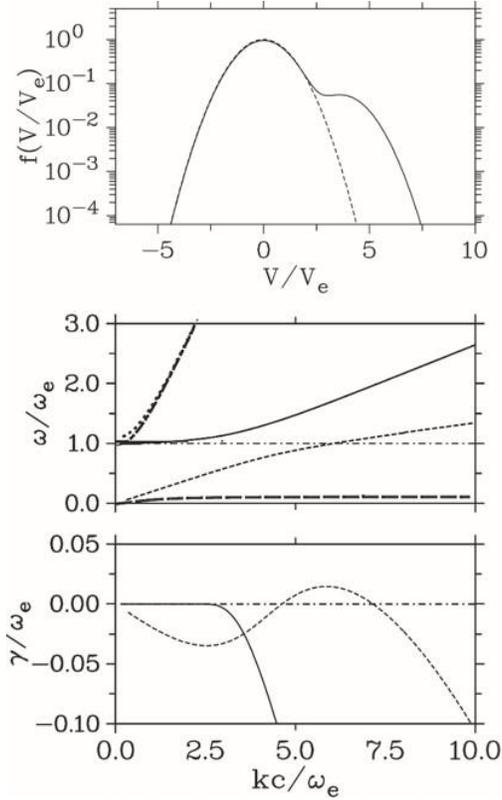

**Figure 1.** Left: a) Electron core-strahl distribution function, b) real and c) imaginary part of $\omega/\omega_e$ versus $kc/\omega_e$. The propagation angle is $\theta=5^0$. The ratio between cyclotron and plasma frequency is $G = \Omega_e/\omega_e = 0.1$.

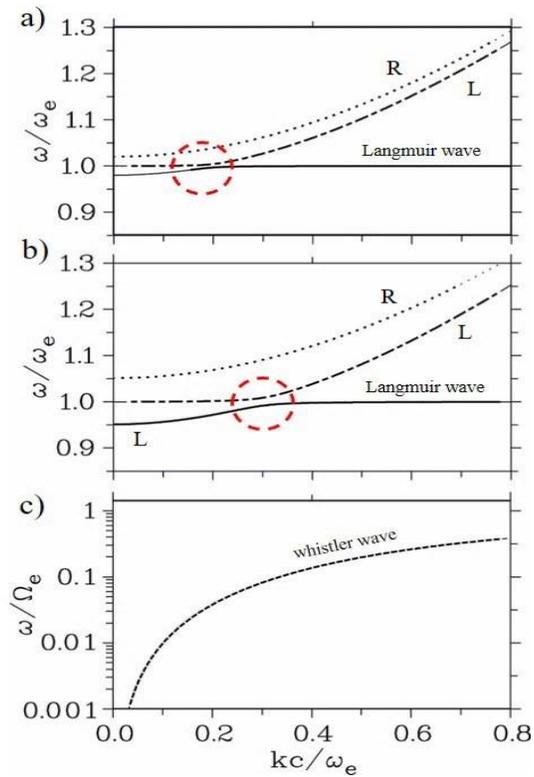

**Figure 2.** Right. Dispersion of the Langmuir wave and adjacent R- and L- electromagnetic modes for θ = 15⁰ and two values of G: a) G = 0.025, b) G = 0.1. In c) the dispersion of the whistler wave is shown which is nearly the same for both values of G. In case of oblique propagation, mode splitting takes place due to the coupling of the Langmuir and the left-handed eletromagnetic wave (L) which cross each other; see the encircled region. With increasing G, the crossing point shifts to larger wave numbers. For G = 0.1, the crossing point is at $kc/\omega_e \sim 0.3$.

**2.2 Short description of the simulation code:**

One-dimensional PIC simulations are performed to investigate the generation of Langmuir oscillations driven by an electron current as well as their conversion into type III radiation and whistler waves as predicted by the fluid model in the companion paper. In these PIC simulations, both electrons and ions are treated as simulation particles (of different masses and signs of charge). The simulations are one-dimensional with the simulation domain along a direction at an angle $\theta = 5°$ with respect to the background magnetic field. Following the wave dispersion analysis described in Section 2.1, the size of the simulation domain is set to be L = 4000$\lambda_e$, where $\lambda_e = c/\omega_e$ is the electron skin length. There are 51200 cells and 1000 simulation particles per cell to represent each particle population. The simulation time step is $\Delta t \omega_e = 3.672 \times 10^{-2}$. The plasma parameters have been chosen same as in the companion paper except that core electrons, strahl electrons (if included), and ions have a finite temperature. The core electrons and ions have Maxwellian distributions of the same temperature, and the corresponding thermal velocity of core electrons is $V_e/c$=0.05. While many simulations have been executed, the results from two representative simulations are presented below in this paper. The first representative simulation (presented in the rest of Section 2) has strahl electrons carrying the initial current, but the other has no strahl electrons and the initial current is carried by core electrons (see Section 3). In the simulation with strahl electrons, the strahl electron density is $n_s/n_0$ = 0.05; strahl electrons initially have a drift Mawellian distribution with the same thermal velocity as core electrons and a drift velocity of $V_s = 4V_e$. This configuration should roughly adapt the electron distribution function which has been measured by Mozer et al. (2024) in connexion with an intensive type III radio burst.

2.3 **Current-driven oscillations of the electron distribution functions:**

According to the fluid theory, the core and halo electrons begin to oscillate under the action of the electric field oscillations which are generated by the current of the strahl. The question was, using kinetic simulations, what effects remain if core and halo are described by Maxwellian distributions instead of delta functions in the fluid approach. The results of the first representative PIC simulation for a core-strahl plasma are shown in Figure 3. As seen there, velocity oscillations occur similar as in the fluid description whereby this is particularly evident in the distribution function of the strahl electrons, see panel b). In general, it can be stated that the oscillations are only detectable in an early phase of development, $\omega_e t \leq 1500$. At a later stage, scattering effects obviously lead to a broadening of the distribution on the low velocity side, which ultimately means heating. In the total distribution function of core and strahl, the oscillations are more difficult to recognize due to the low velocities, which are significantly lower than the thermal velocity of both components. This is ultimately a question of the driving currents, which in the case of higher values can certainly lead to clear effects of electron heating of the core in the form of halo formation. In this context, it is appropriate to refer to the results of Yoon et al. (2012) on theoretical investigations on Langmuir turbulence and supra-thermal electrons. Further discussion is beyond the scope of our considerations.

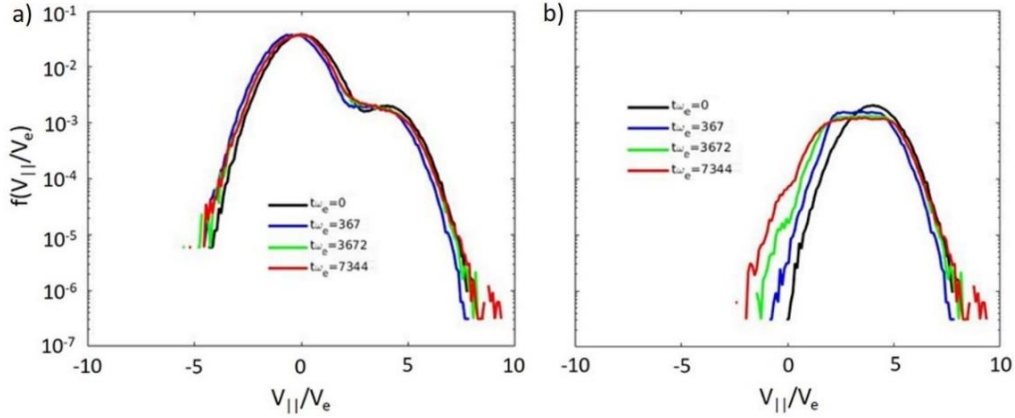

**Figure 3.** Parallel electron distribution function (EDF) at different times in the PIC simulation for a core-strahl plasma. Panel a) represents the total EDF consisting of a Maxwellian for the core and a shifted Maxwellian for the strahl of (relative) density $n_{s0}=0.05$ and a drift velocity of $V_s/V_e=4.0$. An electron thermal velocity of $V_e = 0.05c$ was taken. After a weak beam configuration at the beginning, a plateau is formed which survives up to the final simulation time of $\omega_e t = 7344$.

**2.4 Conversion of strahl-driven Langmuir waves into type III radiation:**
An essential topic of our study concerns the generation of electromagnetic radiation by the coupling of the current-driven Langmuir wave with the intersecting electromagnetic L wave at the crossing point of both wave modes. The basic mechanism of this conversion is already contained in the system of linearized Maxwell-fluid equations. The temporal evolution of the electric and magnetic field components $E_x$, $E_y$, $E_z$, $B_y$ and $B_z$ in the PIC simulation with the strahl electrons are shown in the left panels of Figure 4. For the selected strahl parameters of $n_{s0}=n_s/n_0=0.05$ and $V_{s0} = V_s/c = 0.2$, one gets in agreement with the relation

$$E_{so} = n_{s0}V_{s0}/G \qquad (1)$$

of the fluid theory an initial electric field component of $E_x/E_0 \sim 0.1$ (Figure 4a) where $E_0 = cB_0$. The subsequent decrease up to about $\omega_e t \sim 1500$ characterizes the time interval in which the current-driven Langmuir oscillations lead to the described broadening of the core and strahl distributions. This is followed by the onset of amplitude oscillations with a period of roughly $\omega_e T = 400$ which happen in all electric and magnetic field wave components due to the coupling between the Langmuir wave and the electromagnetic L wave. In the right panels of Figure 4, the spatial profile of $B_z$ and the associated power spectrum are shown. Panel f) exhibits the spatial variation of $B_z$ over the whole simulation box of the length $L=4000\lambda_e$. Its amplitude remains nearly constant at $B_z \sim 0.01B_0$. The high-resolution plot in panel g) indicates the existence of two waves of different wavelengths, which are manifested by two peaks in the corresponding power spectrum, seen in panel h). The first peak belongs to the point of optimum mode coupling at $kc/\omega_e = 0.3$ (see Figure 2b) in complete coincidence with the fluid results. The associated frequency is the electron plasma frequency, $\omega=\omega_e$, marked as F (fundamental). As subsequently shown, the second sharp peak of nearly the same intensity at kc/$\omega_e$ ~ 1.7 is caused by the second harmonic ($\omega = 2\omega_e$), denoted by H.

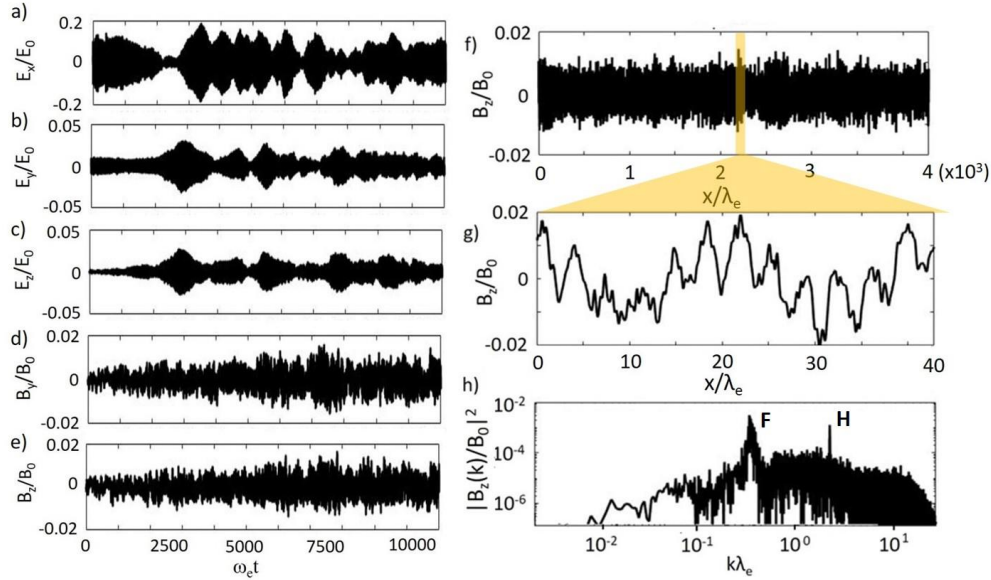

**Figure 4.** (Left) Electric and magnetic field components versus time at x = 2000 $\lambda_e$ in the PIC simulation with the strahl electrons: a) $E_x(t)$, b) $E_y(t)$, c) $E_z(t)$, $E_0=cB_0$, d) $B_y(t)$ and e) $B_z(t)$; (right) spatial profiles of the magnetic field component $B_z$ and the associated wave number power spectrum at $\omega_e t$ = 7000: f) $B_z(x)$ over the whole simulation box, g) $B_z(x)$ in high resolution indicating a superposition of large- and small-scale variations, h) wave number power spectrum $|B_z(k)/B_0|^2$ showing two peaks which belong to the Fundamental F ($\omega = \omega_e$, $k\lambda_e$ ~ 0.3) and the Harmonic H ($\omega = 2\omega_e$, $k\lambda_e$ ~ 1.7).

The hodogram in Figure 5a, in which $E_y$ is shown above $E_x$, also provides valuable information about the temporal development of the electric field. If we compare it with the measurements in Figure 5b of the type III event investigated by Mozer et al. (2024), which served as an orientation with regard to the chosen plasma parameters in the PIC simulation, we find an acceptable correspondence.

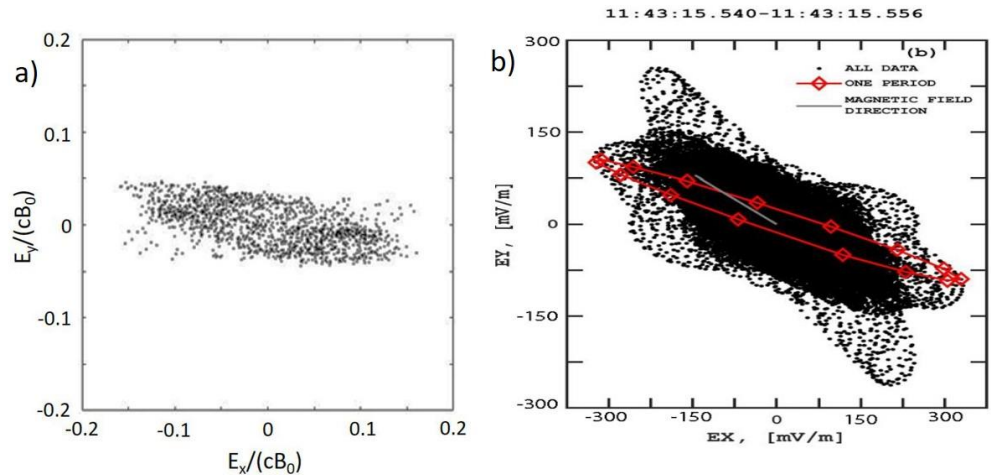

**Figure 5.** Hodogram $E_y$ over $E_x$, a) obtained from the temporal evolution in Figure 4, b) obtained from the electric field measurements by Mozer et al. (2024) during the type III burst on March 21,

2023. The red curve represents one period of the wave. The grey line shows the magnetic field direction.

The temporal evolution of the electric and magnetic field energy in the PIC simulation with the strahl electrons is shown in Figure 6. As seen in panel a), the interval up to $\omega_e t \sim 2000$ is characterized by strong oscillations during which the heating of the core and strahl electrons takes place. This is followed by a slow conversion of electrical energy into magnetic energy. From $\omega_e t \geq 10000$, the transition to a state of saturation becomes apparent.

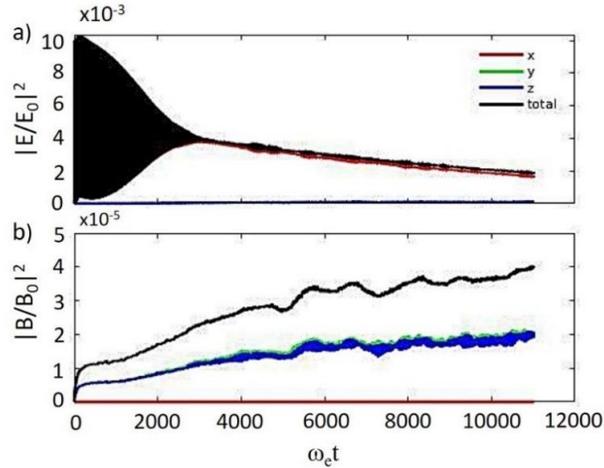

**Figure 6.** Temporal evolution of the electric and magnetic field energy in the PIC simulation with the strahl electrons. a) Electric field energy, b) magnetic field energy. The three components of E and B are marked by red (x), green (y) and blue (z), respectively. The total energy is marked by the black curves.

The frequency spectra of the electromagnetic radiation in the PIC simulation with the strahl electrons, illustrated in Figure 7, provide important information for a comparison with space observations. The bright line in the colour plot of Figure 7a represents the dispersion of the R/L waves as $(\omega/\omega_e)^2 \sim 1+(k\lambda_e)^2$ where F and H mark the first and the second harmonics, respectively. As a remarkable feature one has to note that both emissions, clearly seen in panel b), have nearly the same intensity. That agrees well with the recent observation by Jebaraj et al. (2023) and Chen et al. (2024). Another striking feature is the electromagnetic activity in the frequency range of the electron cyclotron frequency, which is shown in panel c). A maximum exists at $\omega \sim 0.5\Omega_e$ which is a somewhat higher frequency than predicted from the fluid approach of $\omega \sim 0.1\Omega_e$. The question remains open whether the position of the maximum is related to the existence of whistler oscillitons (Sauer et al., 2002), which are characterized by the equality of phase and group velocity.

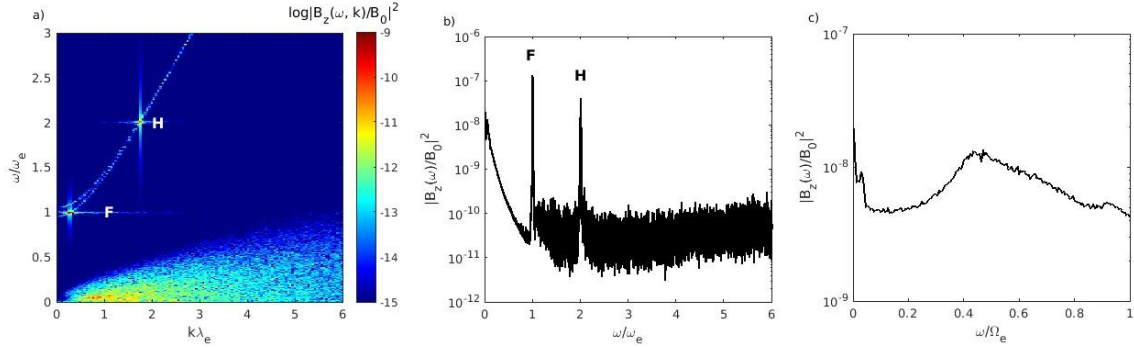

**Figure 7.** Spectra of $B_z$ in the PIC simulation with the strahl electrons. a) Spectrum versus frequency $\omega/\omega_e$ and wavenumber $k\lambda_e$. The Fundamental F ($\omega=\omega_e$, $k\lambda_e\sim0.3$) and the Harmonic H ($\omega=2\omega_e$, $k\lambda_e\sim1.7$) are clearly marked as bright spots. B) Frequency power spectrum showing F and H as pronounced peaks of nearly the same intensity. c) Power spectrum in the frequency range up to $1\Omega_e$. The peak at $\omega \sim 0.5\Omega_e$ indicates the existence of a whistler wave which is simultaneously excited together with the two high-frequency modes.

Finally, Figure 8 shows the power spectra of the three electromagnetic field components $E_x$, $E_z$ and $B_y$ in a high resolution in the wave number range of $0 \leq kc/\omega_e \leq 1$, as never before in kinetic simulations. The mode splitting and the coupling between the current-driven Langmuir wave and the electromagnetic L wave at $kc/\omega_e \sim 0.3$ are clearly seen. That means, the electromagnetic radiation comes from the two modes of mixed polarization which arise at the point where both modes cross each other. The nearly constant amplitude of the Langmuir wave in the $kc/\omega_e \geq 1$ range has important effects on the amplitude of higher harmonics, which is related to the generation of nonlinear currents. Their main contribution comes from the product of the (first order) density and velocity disturbance at the electron plasma frequency which have nearly constant amplitudes in the wave number range of $kc/\omega_e\sim1$. Without going into further details, it should only be mentioned here that this may generate a second harmonic H whose intensity becomes comparable (or even larger) of that of the Fundamental.

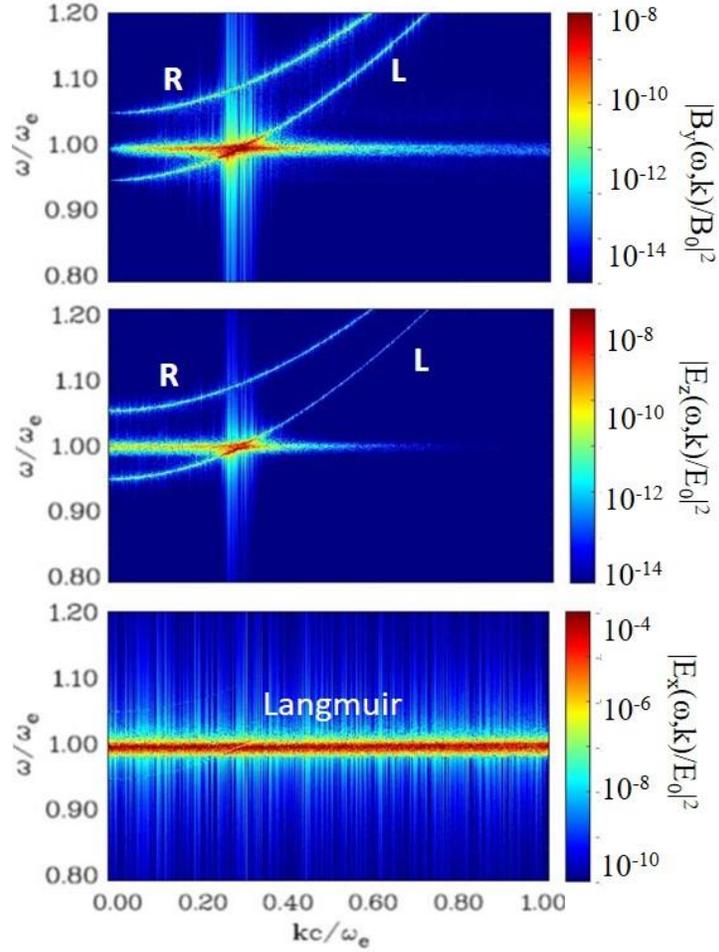

**Figure 8** Power spectra of the electromagnetic field components a) $B_y$, b) $E_z$ and c) $E_x$ in high resolution in the wave number range $0 \leq kc/\omega_e \leq 1$. The hot spots in $B_y$ and $E_z$ at ($\omega \sim \omega_e$, $kc/\omega_e \sim 0.3$) mark the coupling point between the current-driven Langmuir wave and the electromagnetic L wave, see also Figure 2.

## 3 Electromagnetic wave generation by an electron-ion drift

In the PIC simulation presented in this section, in contrast to a strahl producing the current, the sudden onset of a drift of the main electron population relative to the ions is considered. As we will see, this initial configuration leads to the generation of type III radiation and the excitation of whistler waves by the same mechanism as before. However, from a theoretical point of view, the advantage of choosing this situation, which may well be relevant for the interpretation of radiation bursts and whistlers near discontinuities (shocks, boundary layers, etc.), is that generation via beams is excluded from the outset.

### 3.1 Particle oscillations and tail heating
As initial configuration, a velocity shift ($V_d$) of the Maxwellian electron distribution function against the ions of the magnitude of the electron thermal velocity ($V_e$) is assumed, i,e, $V_d = 1V_e$. This means a relatively large value of the initial current, but it allows to get the main effects after a shorter simulation time. The initial configuration and the subsequent evolution of the electron distribution function (EDF) are shown in Figure 9. In panel a), the black curve represents the EDF at t = 0. The further temporal evolution is characterized by oscillations of the EDF whereby at times $\omega_e t \geq 500$ a symmetric heating of the tail electrons occurs, as clearly visible by the widening of the EDF at velocities larger than about three time the thermal velocity. The same can be seen in the colour plot of panel c) which shows the EDF in the $V_\parallel$-$V_\perp$ plane for $\omega_e t = 3672$. The electron oscillations associated with tail heating manifest scattering processes which are interesting with respect to the evolution of the EDF in the young solar wind (e.g., Graham et al., 2017; Cattell et al., 2021; Abraham et al., 2022) and need more detailed studies.

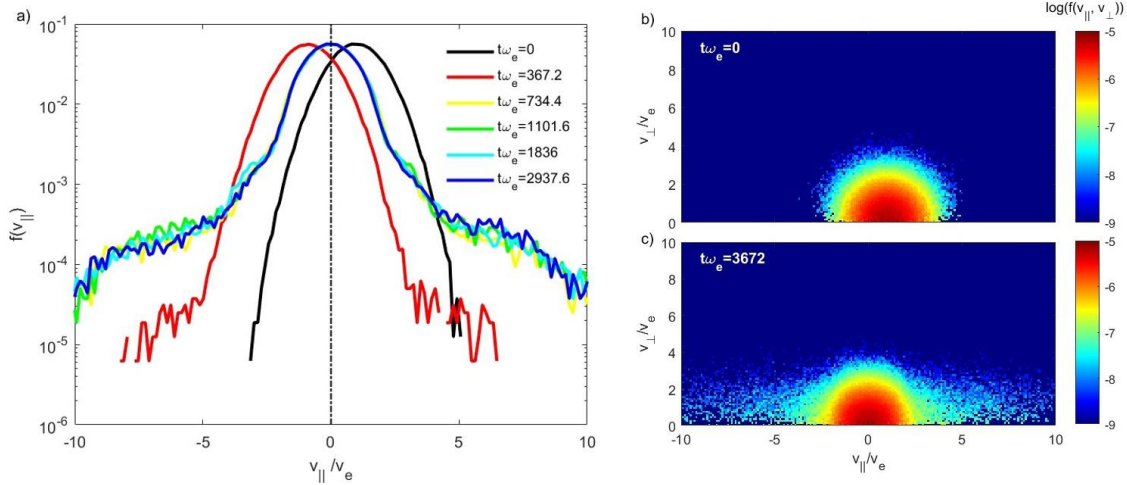

**Figure 9.** a) Temporal evolution of EDF, b) the initial EDF in the $V_\parallel$-$V_\perp$ plane and c) the EDF showing the tail heating at $\omega_e t=3672$ in the PIC simulation without the strahl electrons. The black curve in panel a) shows the initial distribution which is shifted relative to the ions by the velocity $V_d = 1V_e$.

## 3.2 Space-time evolution of the electric and magnetic fields

The temporal development of the electric and magnetic fields driven by an initial electron-ion drift is very similar to that for excitation by a strahl. During the time interval $\omega_e t \leq 500$ (before the tail heating starts), the amplitude of the longitudinal electric field is determined by equation (1) which in the case under consideration (G = 0.1, $V_e/c$ = 0.05) results in an electric field amplitude of $E_x$ = 0.5$E_o$. As can be seen in Figure 10a, this is in good agreement with the result of the simulation. With the onset of tail heating, the amplitude then drops to around $E_x/E_o \sim 0.1$ and remains then largely constant apart from oscillations. As the driving electric field decreases, the magnetic components $B_y$ and $B_z$ in panels d) and e), respectively, build up whereby amplitude oscillations with the period $\omega_e T \sim 500$ become apparent. The associated frequency power spectra are shown in the right panels of Figure 10. Apart from the expected maximum at the electron plasma frequency, the comparable intensity of the fundamental (F) and the second harmonic (H) in the electric and magnetic fields is a remarkable feature.

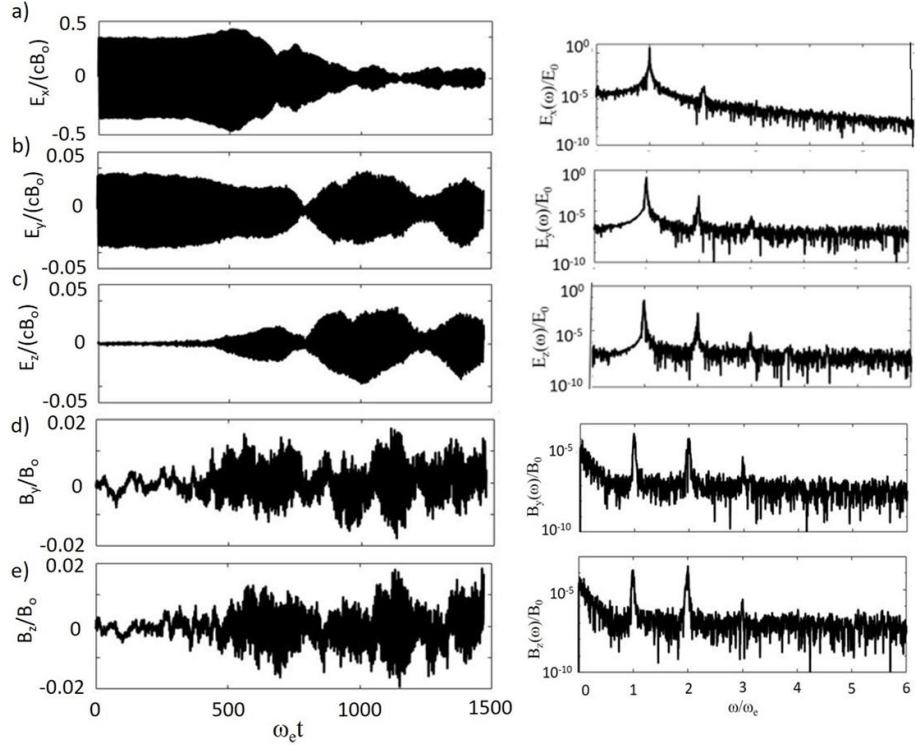

**Figure 10** Left: Temporal evolution of the electric and magnetic fields at x=2000 $\lambda_e$ in the PIC simulation with the core electrons carrying the initial current. a) $E_x$ (t), b) $E_y$(t), c) $E_z$ (t), d) $B_y$(t) and e) $B_z$ (t). Right: Associated frequency power spectra.

Figure 10 uses the magnetic component $B_z$ as example to show how the spatial profiles of the electric and magnetic fields develop over time. Panels a), b) and c) represent $B_z(x)$ over the simulation box of L = 4000$\lambda_e$ at the times $\omega_e t$ = 370, 730 and 3670, respectively. As seen there, the amplitude of $B_z$ first increases with time and then reaches a saturation value of $B_z/B_o$ ~ 0.02. The temporal development of the associated spatial spectrum is very revealing. While at $\omega_e t$ = 370 (right panel a) there is still a broad distribution, a little later at $\omega_e t$ = 730 (right panel b) two clearly pronounced maxima exist, which – absolutely similar to Figure 4h - can be assigned to the fundamental (F) and the second harmonic (H). The fundamental F at $kc/\omega_e$ ~ 0.3 belongs to the crossing point, as marked in Figure 2b, where the Langmuir mode and the L wave intersect. The second harmonic (H) appears according to the dispersion of the (R-, L-) light modes at $kc/\omega_e$ ~ 1.7. The property that the fundamental and the second harmonic have almost the same amplitudes, as seen in the spectra of Figures 10 and 11, is also evident when looking at the spatial profile of $B_z$ with higher resolution (Figure 12). The structure shown in Figure 12 is obviously the result of the superposition of two waves of different wave numbers with comparable amplitudes.

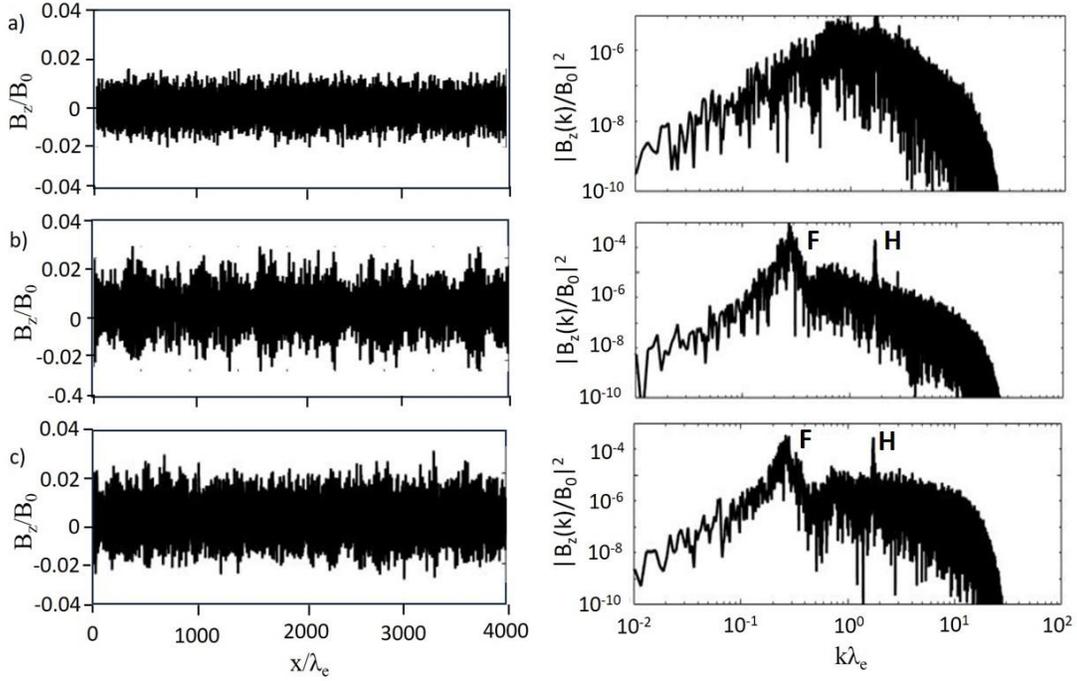

**Figure 11.** Spatial profile of the magnetic component $B_z$ (left) and associated wave number power spectra at three time: a) $\omega_e t = 370$, b) $\omega_e t = 730$, and c) $\omega_e t = 3670$ in the PIC simulation with the core electrons carrying the initial current.

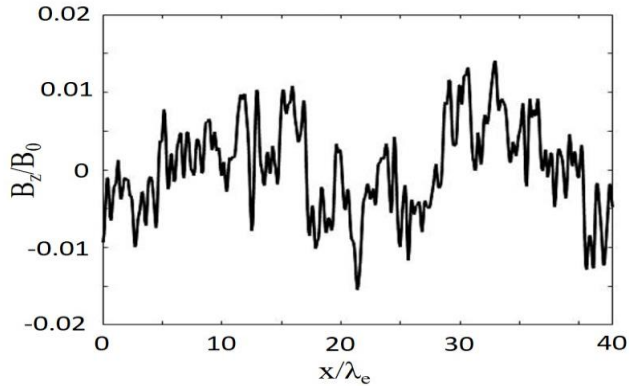

**Figure 12.** Spatial profile of Bz at higher resolution in the PIC simulation with the core electrons carrying the initial current, showing the amplitude oscillation caused by the superposition of the fundamental (F) and the second harmonic (H) of nearly equal amplitudes and wave numbers $kc/\omega_e \sim 0.3$ and $kc/\omega_e \sim 1.7$, respectively. See also Figure 4g for comparison.

### 3.3 Generation of type III radiation and whistler waves

At the end, the main features of the electromagnetic wave generation by a current due to a electron-ion drift are put together. The colour plot of $B_z(\omega,k)$ in Figure 13a shows the both spots which indicate the fundamental (F) and second harmonic (H) emissions. It is clearly seen that the F and H emissions are of nearly the same amplitude. The bright line marks the dispersion of the L- and

R- (light) waves. In the bottom of the ω-k plane one recognizes the whistler wave activity. The increase in intensity in the low frequency range is due to the excitation of whistler waves. In panel c) it can be seen that this has a maximum at ω ~ $0.3\Omega_e$, similar to the triggering by the strahl in Figure 7c. Also noteworthy is the appearance of a weak line at the third harmonic in the frequency spectrum of Figute 13b.

Finally, it should be emphasized once again that no beam-generated Langmuir waves or other instabilities are involved in the creation of either type of electromagnetic wave. They are solely the product of a current that acts as a trigger and can itself have very different origins. In this fundamental sense, the results of the two PIC simulations presented are in agreement with those from the fluid which are presented in the companion paper (Sauer and Liu, 2025).

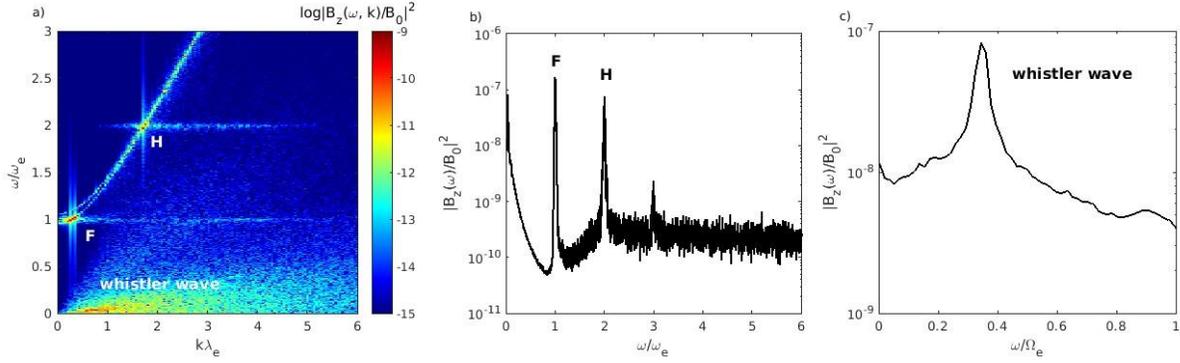

**Figure 13.** Spectra of $B_z$ in the PIC simulation with the core electrons carrying the initial current. a) $|B_z(\omega,k)|^2$ indicating the fundamental (F), the second harmonic (H) and the whistler wave; b) frequency power spectrum $|B_z(\omega)|^2$ in the interval $0 \leq \omega \leq 6\omega_e$; c) power spectrum in the range of whistler waves, $\omega \leq 1\Omega_e$.

## 4 Summary and Discussion

The aim of this work was to find out to what extent basic elements of a simple fluid model for the current-driven generation of electromagnetic waves, worked out in the companion paper (Sauer and Liu, 2025), are confirmed by PIC simulations. This ultimately leads us to the conclusion that the two-step process involving beam-generated Langmuir waves and subsequent wave coalescence according to the ideas of Ginzburg and Zhelezniakov (1958) is not an accurate model to explain type III radiation. Using the linearized system of fluid-Maxwell equations of a core-strahl plasma in the approximation of cold electrons, it has been shown that Langmuir oscillations at the electron plasma frequency are driven by the onset of the strahl current. Moreover, they manifest as velocity oscillations of both the core and strahl electrons. At oblique (quasi-parallel) propagation, the coupling of the Langmuir mode with the electromagnetic L mode leads to amplitude oscillations of all electric and magnetic field components. The wave number of optimum coupling depends on the ratio $G = \Omega_e/\omega_e$ and is given by $kc/\omega_e \sim [G/(1+G)]^{1/2}$. In addition to the electromagnetic waves near $\omega_e$, a whistler wave is driven whose frequency is determined by the wave number of optimum coupling leading roughly to $\omega/\Omega_e \sim G$.

It can be stated that the basic fluid mechanism was confirmed in the kinetic (PIC) simulations. By choosing a sufficiently low beam velocity to suppress the beam instability, the arising oscillations of both the core and the beam electrons are solely caused by the initial current of the strahl. The amplitude of the corresponding electrostatic field is proportional to it and is explicitly given by Equation (1). Furthermore, the wave number spectra of the simultaneously generated electric and magnetic field components show a maximum at the predicted point of coupling between the Langmuir wave and the electromagnetic L wave, that means, at $kc/\omega e \sim 0.3$ for the chosen $G = 0.1$. In addition to the fluid approach which was linear, the second harmonic (H) of nearly the same intensity as the fundamental (F) has been obtained in the simulations. This is a remarkable feature which agrees well with the recent observation by Jebaraj et al. (2023) and Chen et al. (2024).

The second harmonic (H) is particularly suitable for analyzing models of its generation. Nearly identical intensity to the fundamental is difficult to reconcile with the classical concept of coalescence of a beam-generated Langmuir wave with its backscatter counterpart. Surprisingly, comparable intensities of F (or Z mode) and H have been observed in a variety of very different simulations, e.g., in Rhee et al. (2009), Thurgood and Tsiklauri (2015), Chen et al. (2022), and Zhang et al. (2022). Generation of H emission via L+L' interaction also appears questionable given that its intensity remains nearly constant at times when the driving Langmuir waves are already characterized by a significant decrease due to kinetic damping. According to our simulations, in which beam-generated Landau waves generally do not occur, the second harmonic can obviously only arise through nonlinear processes in the wavenumber range $kc/\omega_e \sim 1$. We believe that nonlinearities, which can be triggered by the oscillations of the electron distribution function, shown in Figures 3 and 9, and related nonlinear currents must be considered, but this requires further investigation.

In the second run, as initial configuration a displacement of the whole electron population relative to the ions has been studied. With respect to the conversion of Langmuir oscillations into electromagnetic waves, the results look very similar to the case of using a core-strahl electron distribution. Remarkable, however, is the significant heating of the electrons in the high-energy range. This is obviously caused by scattering of the electrons oscillating in the electric field, which ultimately leads to a distribution consisting of core and halo. The extent to which these results can be of significance for the interpretation of the radial development of the strahl and halo in the solar wind (e.g., Abraham et al., 2022) requires further targeted kinetic simulations. Noteworthy is also the appearance of a weak line at the third harmonic in the frequency spectrum of Figure 13b. Corresponding evidence in interplanetary type II and type III radio bursts is reported by Zlotnik et al. (1998) and Reiner and McDowall (2019), repectively.

The detection of whistler waves in the presented PIC simulations, in which none of the usual instabilities (as temperature anisotropy) are involved, is an important result with regard to the interpretation of a large number of observations in the solar wind, especially in the near solar region by PSP. For a more detailed discussion, please refer to the end of the companion paper on the fluid approach. With regard to the observation of oblique whistler waves, some of which are accompanied by wave numbers of $kc/\omega_e \sim 1$, it will be necessary in the future to carry out corresponding PIC simulations in addition to the results of the fluid approach. It will also be of interest to what extent the intensity of the radiation at fundamental and harmonic frequencies differs from the results for quasi-parallel propagation.

The scattering of strahl electrons by whistler waves is a further much-discussed topic in the literature, e,g., Cattell et al. (2021) and Jagarlamudi et al. (2021). It is generally assumed that these waves, which are generated by kinetic instabilities, cause the scattering. Our picture which is based on our results of the fluid approach and the presented PIC simulations is different. According to this, the particle oscillations that eventually lead to electron heating and the occurrence of whistler waves are an intrinsic expression of the mechanism of current-driven waves in the electromagnetically coupled wave-particle system.

As mentioned already, the main focus of the PIC simulations was restricted to conditions that only the generation of electromagnetic waves in the wave number range $kc/\omega_e < 1$ came into consideration. This was realized by choosing the parameters of the strahl (density, velocity and temperature) in such a way that the formation of an electron-acoustic wave is suppressed by sufficiently large kinetic damping. In further kinetic simulations, the task is thus to abandon the restriction and include effects of mode coupling that take place in the wave range of $k\lambda_D < 1$ with the participation of Langmuir and electron-acoustic waves.

We will conclude our discussion with an outlook on pending simulations. There is no doubt that electron beams and the beam instabilities they cause are directly related. From our studies we suggest that the oscillating electron current because of the instability triggers Langmuir waves in the wavenumber range $k\lambda_e = kc/\omega_e < 1$, which generate electromagnetic radiation through direct transformation. In contrast, previous theoretical investigations and particle simulations have focused on the excitation of Langmuir, backscatter, and ion-acoustic waves with $k\lambda_D < 1$ and their coalescence according to the model of Ginzburg and Zhelezniakov (1958). It is now necessary to analyze the processes occurring in both wavenumber ranges using appropriate particle simulations with sufficient wavenumber resolution. The presented results could be helpful in this regard.